\documentclass[preprint2]{aastex}

\def\ltsima{$\; \buildrel < \over \sim \;$}
\def\simlt{\lower.5ex\hbox{\ltsima}} 
\def\gtsima{$\; \buildrel > \over \sim \;$}
\def\simgt{\lower.5ex\hbox{\gtsima}} 

\def\chandra{{\it Chandra}}

\def\hst{{\it HST}}

\def\merlin{{\it MERLIN}} 

\def\flux{erg cm$^{-2}$ s$^{-1}$}
\def\nh{cm$^{-2}$}
\def\arcsec{$^{\prime\prime}$}

\begin{document}

\title{Detection of an X-ray jet in 3C~371 with \chandra} 

\author{Joseph E. Pesce} \affil{George Mason University, Dept. of
Physics \& Astronomy, 4400 University Dr., M/S 3F3, Fairfax, VA,
22030-4444, and}
\affil{Eureka Scientific, Inc.}

\author{Rita M. Sambruna}
\affil{George Mason
University, Dept. of Physics \& Astronomy and School of Computational
Sciences, 4400 University Dr., M/S 3F3, Fairfax, VA, 22030-4444} 

\author{F. Tavecchio and L. Maraschi}
\affil{Osservatorio Astronomico di Brera, via Brera 28, 20121 Milano, Italy} 

\author{C. C. Cheung}
\affil{Physics Department, MS 057, Brandeis University, Waltham, MA
02454} 

\author{C. Megan Urry}
\affil{STScI, 3700 San Martin Dr., Baltimore, MD 21218}

\author{R. Scarpa}
\affil{European Southern Observatory, 3107 Slonso de Cordova,
Santiago, Chile} 



\begin{abstract}
We report the detection at X-rays of the radio/optical jet of 3C~371,
from a short (10 ks) \chandra\ exposure in March 2000. We also present
a new \merlin\ observation at 1.4 GHz together with a renalysis of the
archival \hst\ WFPC2 F555W image. Despite the limited signal-to-noise
ratio of the \chandra\ data, the X-ray morphology is clearly different
from that of the radio/optical emission, with the brightest X-ray knot
at 1.7\arcsec\ from the nucleus and little X-ray emission from the
brightest radio/optical knot at 3.1\arcsec.
We construct the spectral energy distributions for the two emission
regions at 1.7\arcsec\ and 3.1\arcsec. Both show that the X-ray flux
is below the extrapolation from the radio-to-optical continuum,
suggesting moderately beamed synchrotron from an electron population
with decreasing high energy cut-off as a plausible emission mechanism.
\end{abstract}  


\section{Introduction}


Jets are relatively common in extragalactic radio sources (e.g.,
Bridle \& Perley 1984), and undoubtedly represent a trace of the
energy transport from the nucleus to the outer hot spots/radio lobes.
Hence the interest of observations of jets at shorter wavelengths
which can probe the sites of high energy particle acceleration and
help determine their physical parameters.
About a dozen jets have been detected so far with \hst\
\footnote[2]{Based on observations made with the NASA/ESA {\it Hubble
Space Telescope}, obtained at the Space Telescope Science Institute,
which is operated by the Association of Universities for Research in
Astronomy, Inc., under NASA contract NAS5--26555.} and ground-based
telescopes (Sparks, Biretta, \& Macchetto 1994).  X-ray counterparts,
however, were detected so far in only a handful of cases, including
M87, 3C~273, and Cen~A (Feigelson et al. 1981; Harris \& Stern 1987;
Biretta et al. 1991). This situation is rapidly changing now that the
\chandra\ X-ray Observatory has enabled the discovery of new X-ray
jets (e.g., Chartas et al. 2000; Wilson, Young, \& Shopbell 2001),
thanks to its excellent sensitivity in the 0.2--10 keV range and
superb angular resolution (0.5\arcsec/pixel). Here we report the
detection with \chandra\ of the X-ray counterpart of the radio-optical
jet in the nearby radio source 3C~371 ($z$=0.051), which suggests that
X-ray emission is indeed a common property of such jets.

3C~371 is classified as an intermediate source between BL Lacertae
objects and radio galaxies (Miller 1975). Its radio morphology is
characterized by two giant lobes and a 25\arcsec-long, one-sided jet
(Wrobel \& Lind 1990), typical of FRII galaxies.  The
jet-to-counterjet ratio from the radio is high, 1700:1, implying a
viewing angle $\theta$ \simlt 18$^{\circ}$ and Lorentz factor $\Gamma$
\simgt 3.2 (Gomez \& Marscher 2000). However the lobe separation and
intermediate classification suggest that 3C~371 is most likely seen at
moderate viewing angles, near the upper limit quoted above.

The radio jet is well studied at milli-arcsecond resolution (Gomez \&
Marscher 2000 and references therein) and shows no superluminal
motion, with an upper limit of $\sim 1.4h^{-1}c$. Optical emission
from the jet was detected in ground-based (Nilsson et al. 1997) and in
\hst\ observations (Scarpa et al. 1999). The optical jet is 5\arcsec\
long, with at least three knots from the nucleus; at 1.7\arcsec\ (knot
B, as in Scarpa et al. 1999), 3.1\arcsec\ (knot A, the brightest), and
4.5\arcsec\ (knot D, the faintest). Overall, the optical morphology
tracks the radio one closely in the region of overlap (Scarpa et
al. 1999). 

As part of a \chandra\ AO1 program to image the extended X-ray
environment of BL Lac objects in X-rays (Pesce et al. 2001, in prep.),
we obtained a short (10 ks) exposure of 3C~371 with \chandra,
resulting in a detection of the jet at X-rays. Here we present the
\chandra\ observations, together with unpublished archival \merlin\
1.4 GHz data and a re-extraction of the fluxes from the \hst\ F555W
image (Scarpa et al 1999).  Throughout this work we adopt H$_0=75$ km
s$^{-1}$ Mpc$^{-1}$ and $q_0=0.5$, so at the distance of 3C~371
1\arcsec=0.9~kpc. Preliminary analysis of the \chandra\ observation of
3C~371 was presented by Sambruna et al. (2001a).

\section{Observations and Data Analysis} 

\chandra\ observed 3C~371 on 21 March 2000.  The source was at the
nominal aimpoint of the ACIS-S3 chip and the exposure was continuous
for $\sim$ 10 ks. We used data reprocessed by the \chandra\ X-ray
Center using up-to-date calibration files and standard screening
procedures which left a net exposure of 10,120 s on-source.

As we were originally interested in the large-scale X-ray environment
of the source, no precaution was taken to mitigate pileup of the
central core using a subarray mode.  Thus, the bright central core is
severely piled-up, preventing us from performing any meaningful
spectral and detailed spatial analysis of the nucleus.  An image was
produced by smoothing the raw \chandra\ image with a Gaussian of
width=0.3\arcsec\ in the energy range 0.4--8 keV, where the background
is negligible, with final resolution of 0.86\arcsec\ FWHM. An
elongation of the emission in the S-W direction shows up clearly both
in the raw and smoothed images.  While we can not exclude some
contamination from the PSF wings in the innermost (\simlt 2\arcsec)
parts of the jet, we note that the X-ray jet extends up to $\sim$
4\arcsec. 
More importantly, the X-ray elongation lies at the same position angle
as the known radio-optical jet in 3C~371 (Figure 1), which is
convincing evidence of its reality. 

We also present a new \merlin\ image of 3C~371, calibrated from
unpublished archival data obtained on 1-2 April 1998 as an HI
absorption experiment (P.I., A. Pedlar). The target was observed
continuously for 18 hours, interleaved with 80-second scans of the
nearby phase calibrator 1823+689 every ten minutes.  Some editing and
the initial data calibration were performed at Jodrell Bank which
yielded a single continuum data set at a center frequency of 1.353 GHz
with 7.5 MHz total bandwidth. The absolute flux density was brought to
the Baars et al. (1977) scale with a single 25-minute scan of 3C~286
set to 15.085 Jy. Phase and gain solutions were then applied to 3C~371
using a point source model for 1823+689. The data were edited and
self-calibrated using the Caltech \verb+DIFMAP+ package (Shepherd et
al. 1994).  Appropriate weights were applied to each antenna in order
to account for their differing sensitivities. We reached a dynamic
range (peak to off-source RMS) of about 6000:1 in the final image
(Figure 1) which is not shown at full resolution in order to compare
directly with the \chandra\ image.  Previous 1.7 GHz MERLIN maps shown
at full resolution (Browne \& Orr 1981; Akujor et al. 1994) do not
clearly show knot B, but we clearly detect this feature in our new
image. 

Our \hst\ WFPC2 F555W image was gaussian smoothed
($\sigma\sim$0.21\arcsec) to mirror the \chandra\ resolution and was
presented at full resolution by Scarpa et al. (1999) to which we refer
the reader for details on the calibration.  The F555W filter closely
matches the standard Johnson V-band. Here, we re-extracted fluxes for
knots A and B using apertures of radius 0.75\arcsec\ centered on the
knot positions, for consistency with the \chandra\ extraction.

\section{Results}

Figure 1 shows the 0.4--8 keV ACIS image of 3C~371 (top). Also shown
for comparison on the same scale are the images of the jet in the
optical (middle) and radio (bottom), from our \hst\ and \merlin\
data. The X-ray jet is clearly apparent in the S-W direction, with the
same position angle on the sky as the optical and radio jets. A bright
X-ray structure is visible at $\sim$ 1.7\arcsec\ from the core, after
which the X-ray suface brightness decreases with fainter X-ray
emission present up to $\sim$ 4\arcsec.  A total of 360 counts are
collected from the X-ray jet in 0.4--8 keV, of which $\sim$ 260 are in
the first feature. Due to the limited statistics, it is difficult to
determine whether the first bright ``knot'' is resolved or not.


Comparing the multiwavelength morphologies of the jet, the brighter
structure in the X-ray jet appears to coincide with optical knot B
(following the nomenclature of Scarpa et al. 1999). At this position
both the radio and optical jets show relatively fainter emission
spots. If the X-ray feature is real, this indicates a large difference
of the X-ray morphology with respect to the optical and radio one.
Our optimally designed \chandra\ GO3 observations will better define
the jet X-ray morphology close to the nucleus.

We extracted the X-ray spectrum of knot B using a circular region of
radius 0.75\arcsec. To correct for contamination due to the core PSF,
we extracted spectra from a region with the same radius and at the
same distance from the core (1.8\arcsec), but at different azimuth
positions, to check for any systematic effects. We then subtracted
these spectra from the X-ray spectrum of knot B, and performed
spectral fits on the subtracted data. We find that the spectral fit
results do not depend on the azimuth position.  After subtraction, the
spectrum of knot B contains 229 counts in 0.2--8 keV, and is
consistent with a power law, absorbed by Galactic N$_H$ only ($4.9
\times 10^{20}$ \nh). The column density was fixed to Galactic during
the fits. The fitted photon index is $\Gamma=1.7$ in the range (1.5,
2.1) at 90\% confidence level. The 2--10 keV flux is F$_{2-10~keV}
\sim 1 \times 10^{-13}$ \flux. The fit is good, with $\chi^2$=6 for 9
degrees of freedom. 

For the remaining, fainter part of X-ray jet, we measure $\sim$ 100
counts. Assuming a power-law spectrum with $\Gamma=2$ and Galactic
N$_H$, the 2--10 keV flux is $2.6 \times 10^{-14}$ \flux. This region
coincides with knot A in the optical and radio images (Scarpa et
al. 1999).

\section{Discussion and Conclusions}

Using a short (10 ks) \chandra\ image we have detected the X-ray
counterpart of the radio-optical jet in the nearby radio galaxy
3C~371. Comparison with the optical and radio morphology from a
reanalysis of archival \hst\ and \merlin\ data shows that the peak
X-ray emission from the jet does not coincide with the brightest knot
at longer wavelengths. Future deeper \chandra\ images are needed to
confirm this result. 

Figure 2 shows the Spectral Energy Distributions (SEDs) from radio to
X-rays of knots B and A. For the \merlin\ image, we summed the clean
components within the specified apertures within the (u,v) plane
rather than the image plane.  In both knots, the fluxes at all three
wavelengths were extracted in the same circular region with radius
0.75\arcsec. Figure 2 shows that in both cases the X-ray flux is below
the extrapolation from the radio-to-optical continuum.  The latter is
slightly steeper for knot A than for knot B. 

While the radio to optical continuum is due to synchrotron radiation,
the radiative mechanisms usually considered as the possible origin of
X-rays are either synchrotron or inverse Compton (IC) scattering on
various possible sources of soft photons (Celotti, Ghisellini, \&
Chiaberge 2001).  An analysis similar to PKS 0637--752 (Tavecchio et
al. 2000) indicates that IC emission either on the internal
synchrotron photons (SSC) or on the Cosmic Microwave Background (CMB)
photons can not explain the SED of knot B, unless the source is
characterized by an extremely high Doppler factor ($\delta \sim 50$)
or is very far from equipartition (Figure 3).

The most natural interpretation of the SEDs is that the X-rays are due
to synchrotron emission from the same population of relativistic
electrons responsible for the longer wavelengths. In this scenario the
steepening of the optical-to-X-ray continuum between knots B and A can
be attributed to synchrotron radiative losses. A steep X-ray spectrum
is predicted by this scenario ($\Gamma > 2$) for knot B, consistent
with the fitted value of the photonx index within the 90\%
uncertainties (see above).

Applying a synchrotron homogeneous model, and assuming equipartition,
we derive a magnetic field of few $10^{-5}$ G and a Doppler factor of
a few (for a knot radius of $\sim 10^{21}$ cm, or 1\arcsec),
consistent with the jet-to-counterjet ratio from the radio (\S~1). The
derived electron lifetimes at 1 keV (electrons with Lorentz factor
$\gamma\simeq 3\times 10^7$) are short enough ($\sim 10^{10}$ s) for
the electrons to effectively cool in the knot (Fig. 4). The different
X-ray morphology could be related, in this picture, to the short
cooling time of X-ray emitting electrons.  Electrons producing X-rays
cool rapidly after knot B, while optical and radio electrons can
survive until they reach the bright knot A at $\sim 3$\arcsec, where
emission can be enhanced by compression, by a larger magnetic field
and/or by repeated acceleration with lower maximum energy.

It is instructive to compare 3C~371 with two other well studied
sources with X-ray jets, namely PKS 0637--752 and 3C~273. In the
former, the radio, optical and X-ray morphologies are quite similar
(Chartas et al. 2000; Schwartz et al. 2000) and it is possible to
explain quite well the X-ray emission as due to IC/CMB scattering off
the jet electrons (Tavecchio et al. 2000). For 3C~273 the situation is
more complex, with the global radio and X-ray morphology being
anticorrelated along the jet (Sambruna et al. 2001b). We proposed that
the dominant X-ray emission mechanism is IC from CMB, although the
evidence is not conclusive (see Marshall et al. 2001 for a different
interpretation).  It is tempting to speculate that 3C~371 and PKS
0637--752 represent the prototypes of ``pure'' synchrotron and IC/CMB
sources, respectively, while 3C~273 is an intermediate case, where
both processes contribute to the total X-ray flux. Clearly, a larger
sample of sources with different morphologies and SEDs is necessary to
discuss the importance of the two mechanisms in different sources,
which we anticipate from our ongoing \chandra/\hst\ GO2 survey of
radio jets.


In conclusion, a new X-ray jet has been detected in 3C~371 in a short
\chandra\ exposure. Despite the limited signal-to-noise ratio of the
X-ray data, an intriguing morphology is apparent, with most of the
X-ray counts coming from a region of relative minimum radio and
optical emission. The broad-band SED at two different locations in the
jet are consistent with a synchrotron origin for the X-rays. Deeper
\chandra\ and \hst\ observations of 3C~371 are planned, which will
allow us to confirm the X-ray morphology of the jet and measure
accurately its spectrum as a function of position along the jet.

\acknowledgements 

We thank the referee, George Chartas, for helpful comments. We also
thank Anita Richards for helpful discussions and for performing the
initial calibration of the archive \merlin\ data, Alan Pedlar for
permission to use this data, and Dan Homan for the \verb+PGPLOT+
script used to produce Figure 1. Support for this work was provided by
NASA grants NAS8--39073 (JEP), NAG5--10073 (RMS), and through grant
NAG5--9327, and grant GO-06363.01-95A from the Space Telescope Science
Institute, which is operated by AURA, Inc., under NASA contract
NAS~5-26555 (CMU, RS, CCC).  FT and LM aknowledge partial financial
support from the Italian MURST-COFIN-2000, from ASI-R-105-00 and from
the European Commission ERBFMRX-CT98-0195. \merlin\ is a UK National
Facility operated by the University of Manchester at Jodrell Bank
Observatory on behalf of PPARC.

\newpage 

\begin{figure}
{\Large See color image: fig1.gif}
\caption{Multiwavelength images of the jet of
3C~371. From top to bottom: \chandra\ ACIS-S (0.4--8 keV), \hst\ WFPC2
F555W, and \merlin\ 1.4 GHz. North is up and East is to the left.  The
colors are shown logarithmically; the core and inner jet in the radio
and optical images were saturated in order to show more detail in the
jet. The X-ray image was smoothed with a Gaussian of width 0.3\arcsec,
achieving final resolution of 0.86\arcsec\ FWHM. In order to show the
inner knot B, the optical image was smoothed with a Gaussian of width
0.21\arcsec, achieving final resolution 0.5\arcsec\ FWHM. The radio
image was restored with a beam of 0.5\arcsec\ FWHM. The feature NE of
the optical core is an artifact due to subtraction of the diffraction
spikes.
}
\end{figure}

\begin{figure}
\plotone{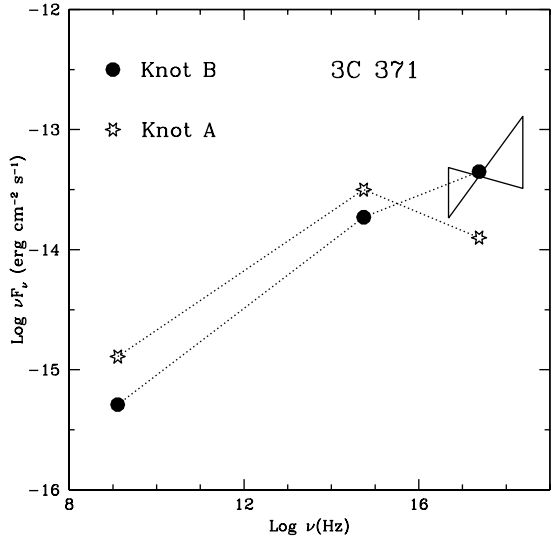}
\caption{Spectral Energy Distributions (SEDs) of two
knots of the 3C~371 jet (see Figure 1 and \S~4). In both cases, the
X-ray flux lies under the extrapolation of the radio-to-optical
continuum. A synchrotron model best accounts for both SEDs (see text),
yielding B=$1.6 \times 10^{-5}$ G, $\delta=6$, $\gamma_{max}=4 \times
10^7$.}
\end{figure}

\begin{figure}
\plotone{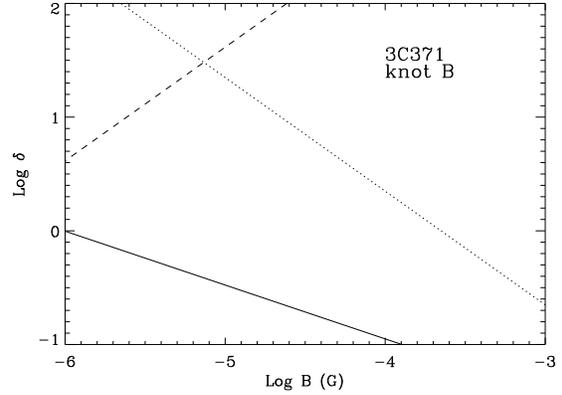}
\caption{Allowed values of the the magnetic field $B$
and the Doppler factor $\delta$ for knot B of 3C~371. {\it Solid
line:} Allowed values of $B, \delta$ if the X-ray emission is due to
SSC.  {\it Dashed line:} Allowed values of $B, \delta$ if the X-ray
emission is due to CMB. {\it Dotted line:} Allowed values under the
assumption of equipartition between the radiating particles and the
magnetic field. Neither SSC or CMB can account for the X-ray emission
from knot B, unless the source is characterized by an extremely high
Doppler factor ($\delta \sim 50$) or is very far from equipartition. 
}
\end{figure}

\begin{figure}
\plotone{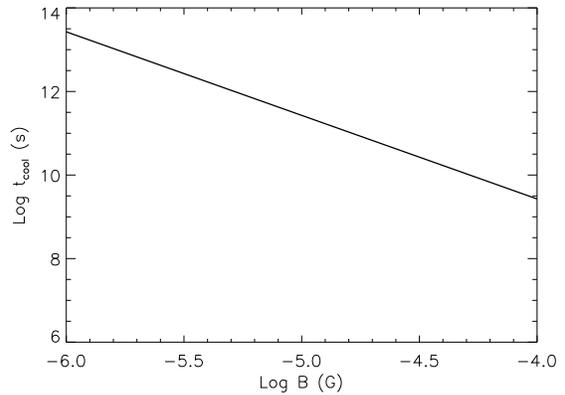}
\caption{Synchrotron cooling time for electrons
producing 1 keV photons for different values of the magnetic field and
assuming equipartition. For $B\simeq 3\times10^{-5}$ G the estimated
cooling time is $\sim 10^{10}$ s (see text).
}
\end{figure}

\end{document}